\documentclass{elsart}
\begin{document}

\begin{frontmatter}
\title{Identifying Complexity by Means of Matrices}

\author{S. Dro\.zd\.z$^{1,2,3,4}$, J. Kwapie\'n$^{1,2}$,
J. Speth$^1$, M. W\'ojcik$^2$}

\address {$^1$Institut f\"ur Kernphysik, Forschungszentrum J\"ulich,
D-52425 J\"ulich, Germany, \\
$^2$Institute of Nuclear Physics, PL--31-342 Krak\'ow, Poland, \\
$^3$Institute of Physics, University of Rzesz\'ow,
PL--35-310 Rzesz\'ow, Poland, \\
$^4$Physikalisches Institut, Universit\"at Bonn, D-53115 Bonn, Germany}

\begin{center}
{\it\small Dedicated to Professor Dr. H.E. Stanley on the occasion
of his 60th birthday}
\end{center}

\begin{abstract}
Complexity is an interdisciplinary concept which, first of all,
addresses the question of how order emerges out of randomness.
For many reasons matrices provide a very practical and powerful tool
in approaching and quantifying the related characteristics.
Based on several natural complex dynamical systems, like the strongly
interacting quantum many-body systems, the human brain and the
financial markets, by relating empirical observations to the random
matrix theory and quantifying deviations in term of a reduced
dimensionality, we present arguments in favour of the statement
that complexity is a pheomenon at the edge between collectivity and chaos.
\end{abstract}

\begin{keyword}
Natural complex systems \sep Random matrix theory \sep Order out of randomness
\PACS 05.30.Fk \sep 21.60.-n \sep 24.60.Lz \sep 21.10.Re
\end{keyword}
\end{frontmatter}

\section{Introduction}

By its very nature, even though central to the contemporary physics,
the concept of complexity still lacks a precise definition.
In qualitative terms this concept refers to diversity of forms,
to emergence of coherent patterns out of randomness and also
to some ability of frequent switching among such patterns.
This normally involves many components, many different space
and time scales, and thus such phenomena like chaos, noise,
but, of course, also collectivity and criticality~\cite{Stanley}.
In fact, due to all those elements, it seems most appropriate
to search for a real complexity just at the interface of chaos
and collectivity~\cite{Kauffman,Bak}. Indeed, these two seemingly
contradictory phenomena have to go in parallel, as they both are
connected with existence of many degrees of freedom 
and a strong, often random, interaction among them.

Approaching complex systems, either empirically or theoretically,
is typically based on analyzing large multivariate ensembles of
parameters. For this reason, probably the most efficient formal 
frame to quantify the whole variety of effects connected with 
complexity is in terms of matrices.
Since complexity is embedded in chaos, or even noise, the random
matrix theory (RMT)~\cite{Wigner,Mehta} provides then an appropriate
reference. Its utility results predominantly from the fact
that the degree of agreement quantifies the generic properties of
a system - those connected with chaotic or noisy activity.
For the complex systems this is expected to be a dominant component,
but this component is not what constitutes an essence of complexity.
From this perspective the deviations are even more relevant and
more interesting as they reflect a creative and perhaps
deterministic potential emerging from a noisy background of such systems.
The main related purpose of the present summary is to identify,
within the matrix formalism, some principal characteristics of such
deviations - the ones that are common and typical to natural complex dynamical
systems.

\section{Coherence versus noise in matrix representation}

Expressed in the most general form, in essentially all the cases 
of practical interest, the $n \times n$ matrices ${\bf W}$ used
to describe the complex system are by construction designed as

\begin{equation}
{\bf W} = {\bf X} {\bf Y}^T,
\label{cab}
\end{equation}
where ${\bf X}$ and ${\bf Y}$ denote the rectangular $n \times m$
matrices. Such, for instance, are the correlation matrices whose 
standard form corresponds to ${\bf Y} = {\bf X}$. In this case
one thinks of $n$ observations or cases, each represented by a $m$
dimensional row vector ${\bf x}_i$ $({\bf y}_i)$,
($i=1,...,n$)~\cite{Muirhead}, and typically $m$ is larger than $n$. 
In the limit of purely random correlations
the matrix ${\bf W}$ is then said to be a Wishart matrix~\cite{Wishart}.
The resulting density ${\rho_{\bf W}(\lambda)}$ of eigenvalues is
here known analytically~\cite{Edelman}, with the limits
$(\lambda_{min} \le \lambda \le \lambda_{max})$ prescribed by
$\lambda^{max}_{min} = 1 + 1/Q \pm 2\sqrt{1/Q}$
and $Q = m/n \ge 1$.
The variance of the elements of ${\bf x}_i$ is here assumed unity.

The more general case, of ${\bf X}$ and ${\bf Y}$ different, results in
asymmetric correlation matrices with complex eigenvalues $\lambda$.
As shown recently~\cite{Kwapien}, such matrices also turn out to 
provide a very powerful tool in practical applications.
In this more general case a limiting distribution corresponding to
purely random correlations seems not to be yet known analytically 
as a function of $m/n$. The result of ref.~\cite{SCSS} indicates however
that in the case of no correlations, quite generically, one may expect
a largely uniform distribution of $\lambda$ bound in an ellipse 
on the complex plane.

Further examples of matrices of similar structure, of great interest
from the point of view of complexity, include the Hamiltonian
matrices of strongly interacting quantum many body systems
such as atomic nuclei. This holds true~\cite{Drozdz1} on the level
of bound states where the problem is described by the Hermitian 
matrices, as well as for excitations embedded in the continuum.
This later case can be formulated in terms of an open quantum 
system~\cite{Rotter}, which is represented by a complex
non-Hermitian Hamiltonian matrix. Several neural network models
also belong to this category of matrix structure~\cite{Hopfield}.
In this domain the reference is provided by the Gaussian (orthogonal,
unitary, symplectic) ensembles of random matrices with the semi-circle
law for the eigenvalue distribution~\cite{Mehta}. For the irreversible
processes there exists their complex version~\cite{Ginibre}
with a special case, the so-called scattering ensemble~\cite{SZ},
which accounts for $S$-matrix unitarity.

As it has already been expressed above, several variants of ensembles
of the random matrices provide an appropriate and natural reference
for quantifying various characteristics of complexity.
The bulk of such characteristics is expected to be consistent
with RMT, and in fact there exists strong evidence that it is.
Once this is established, even more interesting are however deviations,
especially those signaling emergence of synchronous or coherent
patterns, i.e.,
the effects connected with the reduction of dimensionality.
In the matrix terminology such patterns can thus be associated with
a significantly reduced rank $k$ (thus $k \ll n$)
of a leading component of ${\bf W}$.
A satisfactory structure of the matrix that would allow some 
coexistence of chaos or noise and of collectivity thus reads:
\begin{equation}
{\bf W} = {\bf W}_r + {\bf W}_c.
\label{ccc}
\end{equation}
Of course, in the absence of ${\bf W}_r$, the second term $({\bf W}_c)$
of ${\bf W}$
generates $k$ nonzero eigenvalues, and all the remaining ones $(n-k)$
constitute the zero modes. When ${\bf W}_r$ enters as a noise
(random like matrix) correction, a trace of the above effect is
expected to remain, i.e., $k$ large eigenvalues and the bulk composed
of $n-k$ small eigenvalues whose distribution and fluctuations are
consistent with an appropriate version of random matrix ensemble.
One likely mechanism that may lead to such a segregation of
eigenspectra is that $m$ in eq.~(\ref{cab}) is significantly smaller than
$n$, or that the number of large components makes it effectively
small on the level of large entries $w$ of ${\bf W}$.
Such an effective reduction of $m$ $(M=m_{eff})$ is then expressed by
the following distribution $P(w)$ of the large off-diagonal matrix
elements in the case they are still generated by the random
like processes~\cite{Drozdz1}:
\begin{equation}
P(w) = { { \vert w \vert^{(M-1)/2} K_{(M-1)/2} (\vert w \vert) } \over
{ 2^{(M-1)/2} \Gamma(M/2) \sqrt{\pi} }},
\label{eqPr}
\end{equation}
where $K$ stands for the modified Bessel function.
Asymptotically, for large $w$, this leads to
$P(w) \sim \exp (-\vert w \vert)~{\vert w \vert}^{M/2-1}$,
and thus reflects an enhanced probability of appearence of a few
large off-diagonal matrix elements as compared to a Gaussian
distribution. As consistent with the central limit theorem
the distribution (\ref{eqPr}) quickly converges to a Gaussian
with increasing $M$.

Another mechanism that may lead to a structure analogous
to (\ref{ccc}), is the presence of some systematic trend,
in addition to noise, in the ${\bf X}$ and ${\bf Y}$ matrices.
Then~\cite{Drozdz2}, to a first approximation, the second term 
in this decomposition is represented just by a matrix whose all entries 
are close in magnitude, and thus its rank is directly seen to be unity.
The most straightforward indication that this kind of decomposition
applies is an asymmetric shift of $P(w)$ relative to zero.

Based on several examples of natural complex dynamical systems, like
the strongly interacting Fermi systems, the human brain and the 
financial markets, below we systematize evidence that such effects
are indeed common to all the phenomena that intuitively can be
qualified as complex.

\section{Common features of complexity in natural systems}

Since it was nuclear physics which gave birth to several concepts
relevant to the physics of complex systems, in particular to RMT,
we begin with an issue which originates from nuclear considerations
and which, at the same time, addresses a problem~\cite{Johnson}
of great current interest, attracting lot of activity in the literature.
More specifically, the related question asks what is a nature of the
ground state if the two-body interaction is drawn from a Gaussian
ensemble. This is an example of a sparser connectivity than just
everything with everything and by this it is more realistic.

In the presence of two-body interactions the many-body Hamiltonian
matrix elements $v^J_{\alpha,\alpha'}$ of good total
angular momentum $J$ in the shell-model basis $\vert {\alpha} \rangle$ 
generated by the mean field, can be expressed as follows~\cite{Talmi}:
\begin{equation}
v^J_{\alpha,\alpha'} = \sum_{J' i i'} c^{J \alpha \alpha'}_{J' i i'}
g^{J'}_{i i'}.
\label{eqv}
\end{equation}   
The summation runs over all combinations of the two-particle states 
$\vert i \rangle$ coupled to the angular momentum $J'$ and connected 
by the two-body interaction $g$. The analogy of this structure
to the one schematically captured by the eq.~(\ref{ccc}) is evident.
$g^{J'}_{i i'}$ denote here
the radial parts of the corresponding two-body matrix elements 
while $c^{J \alpha \alpha'}_{J' i i'}$ 
globally represent elements of the angular momentum recoupling geometry. 
$g^{J'}_{i i'}$ are drawn from a Gaussian distribution while
the geometry expressed by 
$c^{J \alpha \alpha'}_{J' i i'}$ 
enters explicitly. An explicit calculation then shows~\cite{Drozdz1}
that for $J > 0$ the tails of $P(v)$ are very nicely reproduced by
the eq.~(\ref{eqPr}) with $M \approx 2$. This originates
from the fact that a quasi-random coupling of individual spins 
results in the so-called geometric chaoticity~\cite{Ericson} and thus
$c^{J \alpha \alpha'}_{J' i i'}$ coefficients are also Gaussian
distributed. 
In this case, these two ($g$ and $c$) essentially random ingredients
lead however to an order of magnitude larger separation of the ground
state from the remaining states as compared to a pure RMT limit,
and this is consistent with the above estimate for $M$.
Due to more severe selection rules the effect of geometric chaoticity 
does not apply for $J=0$.
As a consequence, in this particular case $P(v)$
is much closer to a Gaussian, i.e., $M$ is here much larger.
Consistently, the ground state energy gaps measured relative to the 
average level spacing characteristic for a given $J$ is larger
for $J > 0$ than for $J = 0$, and also $J > 0$ ground states are
more orderly than those for $J = 0$, as it can be quantified in terms
of the information entropy~\cite{Drozdz2}, for instance.

Interestingly, such reductions of dimensionality of the Hamiltonian
matrix can also be seen~\cite{Drozdz3} locally in explicit calculations 
with realistic (non-random) nuclear interactions.
A collective state, the one which turns out coherent with some operator
representing physical external field, is always surrounded 
by a reduced density of states, i.e., it repells the other states.
It is also appropriate to mention at this point that similar effects 
of reduced dimensionality, applicability of the formula (\ref{eqPr}),
and of the resulting segregation of states one 
observes~\cite{IMR,Drozdz4,DTW} in the many body quantum open
systems due to the coupling to continuum. Of course, in the latter case
on the complex plane. In all those cases, the global fluctuation
characteristics remain however largely consistent with the corresponding
version of the random matrix ensemble.

Recently, a broad arena of applicability of the random matrix theory
opens in connection with the most complex systems known to exist
in the universe. With no doubt, the most complex is the human's
brain and those phenomena that result from its activity.
From the physics point of view the financial world, reflecting such
an activity, is of particular interest~\cite{MS}
because its characteristics are quantified directly in terms of numbers and
a huge amount of electronically stored financial data is readily 
available. An access to a single brain activity is also possible
by detecting the electric or magnetic fields generated by the neuronal
currents~\cite{HHIKL}. With the present day techniques of electro-
or magnetoencephalography, in this way it is possible to generate
the time series which resolve neuronal activity down to the scale of 1 ms.

One may debate over what is more complex, the human brain or the
financial world, and there is no unique answer. It seems however
to us that it is the financial world that is even more complex.
After all, it involves the activity of many human brains and it
seems even less predictable due to more frequent changes between
different modes of action. Noise is of course owerwhelming in either
of these systems, as it can be inferred from the structure 
of eigenspectra of the correlation matrices taken across different
space areas at the same time~\cite{Laloux,Plerou,Drozdz2}, or across
different time intervals~\cite{Kwapien,Drozdz5}.
There however always exist several well identifiable deviations,
which, with help of reference to the universal characteristics of
the random matrix theory, and with the methodology briefly reviewed
above, can be classified as real correlations or collectivity.
An easily identifiable gap between the corresponding eigenvalues
of the correlation matrix and the bulk of its eigenspectrum
plays the central role in this connection.
The brain when responding to the sensory stimulations develops larger
gaps than the brain at rest~\cite{Kwapien}. The correlation matrix
formalism in its most general asymmetric form allows to study~\cite{Kwapien}
also the time-delayed correlations, like the ones between the oposite 
hemispheres. The time-delay reflecting the maximum of correlation
(time needed for an information to be transmitted between the different
sensory areas in the brain~\cite{Kwapien1,ILKDS}) is also associated with
appearance of one significantly larger eigenvalue.
Similar effects appear to govern formation of the heteropolymeric
biomolecules. The ones that nature makes use of are separated by
an energy gap from the purely random sequences~\cite{Shak}.

As far as the dynamics of evolution of complex systems is concerned
one interesting observation made in~\cite{Drozdz2},
based on the stock market evolution, is also to be pointed out
in the present context.
It appears that increases, as a rule, are more competitive, 
and thus less collective than decreses which are always accompanied 
by a more violent collectivity.
This may illustrate a more general logic of evolution of natural
complex dynamical systems.

Such characteristics of coherence are typically connected with
a few largest eigenvalues of the correlation matrix and those
eigenvalues stay significantly above $\lambda_{max}$.
The bulk of eigenvalues is quite universally
consistent with the RMT limit. There exist however some
more subtle measures of eigenvalue fluctuations in the random matrix
theory. In particular, we refer here to the Tracy-Widom law~\cite{Tracy}
which, based on Painlev\'e representations and a proper scaling of
eigenvalues, provides a general formalism to study fluctuations of
individual eigenvalues. An appropriate rescaling of eigenvalues
makes this law applicable also to Wishart matrices~\cite{Johnstone}.
Motivated by this formalism, and based on the high frequency recordings
of all the stocks comprised by DAX, we recently analysed the fluctuations
of various eigenvalues of the correlation matrix using our 
methodology~\cite{Drozdz2} of moving time-window. Even though, on average, the
distribution of the bulk of eigenvalues is here consistent with the
RMT limit, significant deviations relative to fluctuations
of eigenvalues of the correlation matrices calculated from purely random time
series still remain. This can partly be attributed to the effect
reminiscent of the slaving principle of synergetics~\cite{Haken}:
if one, or a small fraction of states take the entire collectivity, all the
others become enslaved and thus their 'noise freedom' gets also reduced.
This may affect the oposite edge of the spectrum by suppressing the amount
of noise there, and thus making the corresponding states again deviating
more from their RMT limit. The above seems to provide a likely explanation
for the effect observed in ref.~\cite{Plerou}, that the smallest eigenvalues
of the financial correlation matrix also correspond to more localised
states.
Such effects seem to constitute another manifestation of complexity.

\section{Summary}

The above brief review tempts to view complexity as a trinity
comprising coherence, chaos and a gap (probably not too large) between them.
Coherence constitutes the essence as it makes patterns and structures,
which is of primary interest and importance.
Chaos is always present in any really interesting system and, in fact,
it is even needed as it allows to quickly explore the whole
available phase space, and thus to probe various possibilities 
and to switch from one pattern of activity to another.
Finally, the gap between them allows the structures to be identifiable 
and to exist for some time. Thus all the three are needed in parallel.
Such a combination probably makes a natural system most efficient 
in its evolution. 

This work was partly supported by KBN Grant No. 2 P03B 097 16 
and by the German-Polish DLR scientific exchange program, 
grant No. POL-028-98.

\newpage

\end{document}